\documentclass{revtex4}
\usepackage{amsmath}
\usepackage{amssymb}
\usepackage{graphicx}
\usepackage{bm}
\usepackage{dcolumn}

\newcommand{\beq}{\begin{equation}}
\newcommand{\eeq}{\end{equation}}

\begin{document}

\title{Electron mobility on a surface of dielectric media: influence of
surface level atoms }
\author{P. D. Grigoriev}
\email{grigorev@itp.ac.ru}
\author{A.M. Dyugaev}
\affiliation{L. D. Landau Institute for Theoretical Physics, Chernogolovka, 
Moscow region, 142432, Russia}
\author{E.V. Lebedeva} 
\affiliation{Institute of Solid State Physics,
Chernogolovka, Moscow region, 142432, Russia}

\begin{abstract}
We calculate the contribution to the electron scattering rate from the
surface level atoms (SLA), proposed in [A.M. Dyugaev, P.D. Grigoriev, 
JETP Lett. 78, 466 (2003)]. The inclusion
of these states into account was sufficient to explain the long-standing
puzzles in the temperature dependence of the surface tension of both He 
isotopes and to reach a very good agreement between theory and experiment. 
We calculate the contribution from these SLA to the surface electron scattering
rate and explain some features in the temperature
dependence of the surface electron mobility. This
contribution is essential at low temperature $T<0.5$ when the
He vapor concentration is exponentially small. 
For an accurate calculation of the electron mobility
one also needs to consider the influence of the clamping electric
field on the surface electron wave function and the temperature dependence
of the He3 chemical potential. 
\end{abstract}

\maketitle

Two-dimensional electron gas on a surface of dielectric media is wide a subject
of research for several decades (for reviews see, e.g., \cite%
{Shikin,Edelman,Monarkha}). The electrons are attracted to the dielectric
interface by electric image forces and become localized in the direction
perpendicular to the surface. The surface of liquid helium has no solid
defects (like impurities, dislocations, etc.) and gives a unique chance to
create an extremely pure 2D electron gas. This system simulates the
solid-state 2D quantum wells without disorder. Many fundamental properties
of a 2D electron gas have been studied with the help of electrons on the
surfaces of liquid helium. The many-body electron effects on the surface of
liquid helium are determined by the interaction between electrons
and surface waves (ripplons) and by the Coulomb electron-electron (e-e)
interaction screened by a substrate. The Wigner crystallization of 2D electron gas,
induced by the Coulomb e-e interaction, was first observed and extensively
studied on the surface of liquid helium (see \cite%
{Shikin,Edelman,Monarkha} for reviews). A large number of interesting quantum objects
may be experimentally realized with electrons on the liquid helium surface:
the quantum rings, the 1D electron wires, the quantum dots etc. The
electrons on the liquid helium surface may also serve for an experimental
realization of a set of quantum bits with very long incoherence time.\cite%
{DykmanQC} All these quantum objects and phenomena depend in a crucial way
on the structure and properties of the liquid helium surface itself.

The interface between liquid helium and vacuum is usually supposed to be
sharp -- the density of helium atoms falls to zero on a distance of
intermolecular spacing which is much smaller than the size of the surface
electron wave function. The electrons are clamped to the surface by the
image forces and by external electric field. The electron wave function on
the lowest energy level in $z$-direction without external field is given by 
\cite{Shikin,Edelman,Monarkha} 
\begin{equation}
\Psi _{e}(z)=2\gamma ^{3/2}z\exp (-\gamma z),  \label{Psiz}
\end{equation}%
where in the absence of clamping electric field $\gamma =m\Lambda /\hbar
^{2} $ and $\Lambda \equiv e^{2}(\varepsilon -1)/4(\varepsilon +1)$. The
dielectric constants of liquid $^{4}He$ is $\varepsilon _{4}=1.0572$ and $%
\gamma =(76$\AA $)^{-1}$, while for $^{3}He$ $\varepsilon _{3}=1.0428$ and $%
\gamma =(101$\AA\ $)^{-1}$. Hence, the electron wave function is rather
extended in $z$-direction, which reduces the influence of small surface
ripple on the electron motion and makes the mobility of 2D electrons gas on
the helium surface to be rather high. At low electron concentration $%
N_{e}\approx 10^{7}1/cm^{2}$ the mobility of electrons on the liquid helium
surface at $T=0.1K$ reaches $10^{4}m^{2}/$V$\sec $.\cite{ExpMobil}, which is
about $10^{4}$ times greater than the carrier mobility in heterostructures.
For several decades people believed that at low enough temperature, i.e.
when the concentration of $He$ vapor is exponentially small, on the surface
of liquid helium there is only one type of excitations interacting with
surface electrons. These excitations are the quanta of surface waves, called
ripplons. These surface waves determine the temperature dependence of the
surface tension of liquid helium, as well as the mobility, cyclotron
resonance line-width and other properties of surface electrons at
temperature below $\sim $0.5K.\cite{Shikin,Edelman,Monarkha}

Recently, we proposed\cite{SurStates} that in addition to the ripplons,
there is another type of surface excitations -- the surface level atoms
(SLA). The SLA considerably influence the temperature dependence of the
surface tension of both liquid helium isotopes. Taking the SLA into account
we explained the long-standing puzzles\cite{SurStates,SurTensExp} in this
temperature dependence and reached a very good agreement (up to 0.1\%)
between theory and experiment on the surface tension in a large temperature
interval. This high agreement may serve as an additional proof of the
existence of SLA as the new type of excitations on the surface of liquid
helium.

An accurate microscopic description of this new type of excitations is a
rather complicated problem that requires an investigation of many-particle
interactions in liquid helium. However, one may consider SLA
phenomenologically, assuming them to be similar to the quantum states of
helium atoms which are localized at the liquid helium surface. The SLA may
also propagate in the surface plane and have the quadratic dispersion%
\begin{equation*}
\varepsilon (k)=E_{SLA}+k^{2}/2M^{\ast },
\end{equation*}%
where $k$ is the 2D momentum of this excitation along the surface. Both the
SLA energy $E_{SLA}$ and the effective mass $M^{\ast }$ of the SLA differ
for the two He isotopes $^{3}$He and $^{4}$He. The SLA energies $E_{SLA}$
are intermediate between the energy of He atom in vacuum $E_{vac}$\ and the
chemical potential $\mu $ of this atom inside the liquid. If one takes the
energies of He atoms in vacuum to be zero, $E_{vac}^{He}=0$, the chemical
potential at $T\rightarrow 0$ is $\mu ^{He4}=-7.17K$ and $\mu ^{He3}=-2.5K$,
while the energies of these atom on the surface levels, as suggested by the
temperature dependence of the surface tension,\cite{SurStates} are $%
E_{SLA}^{He4}\approx -3.5K$ and $E_{SLA}^{He3}\approx -2.25K$. Therefore, at
low enough temperature the concentration of SLA becomes exponentially higher
than the He vapor concentration, and the influence of the SLA on the
properties of surface electrons becomes more important than the influence of
the He vapor.

On the other hand, there is a long-standing discrepancy between the
theoretical prediction\cite{Saitoh} for the electron mobility on the liquid
helium surface and the experimental data (see Fig. 2 of Ref. \cite{ExpMobil}%
). According to the existing theory\cite{Saitoh} the ratio of electron
mobilities on $^{3}$He and $^{4}$He surfaces at the same concentration of
helium vapor must be equal to $\gamma _{He4}/\gamma _{He3}=1.33$, while the
experimental lines (see Fig. 2 of Ref. \cite{ExpMobil}) even cross each
other. In the present report we calculate the influence of the SLA on the
mobility of 2D surface electrons and analyze if taking into account this
influence helps to explain the existing discrepancy between the theory and
experiment. Note that the similar surface states may also occur in other
liquids and solids, as solid hydrogen or neon, raising similar questions.

Vapor atoms or SLA can be considered as point-like impurities localized at
points $r_{i}$. These impurities interact with electrons via a $\delta $%
-function potential $V_{i}(r)=U\delta \left( r-r_{i}\right) $. Then there is
no difference between transport and usual mean free time $\tau $, which is
given by:%
\begin{equation}
\frac{1}{\tau }=\frac{2\pi }{\hbar }\int dzN_{He}^{tot}(T,z)\int \frac{d^{2}%
\mathbf{p}^{\prime }\left\vert T_{\mathbf{pp}^{\prime }}\left( z\right)
\right\vert ^{2}}{(2\pi \hbar )^{2}}\delta \left( \varepsilon _{\mathbf{p}%
}-\varepsilon _{\mathbf{p}^{\prime }}\right) ,  \label{tau}
\end{equation}%
where $\epsilon (\mathbf{p})=\mathbf{p}^{2}/2m^{\ast }$ is electron
dispersion relation, $m^{\ast }$ is the effective electron mass, $\left\vert
v_{\mathbf{p}}\right\vert =p/m^{\ast }$ is the electron velocity. The 2D
matrix element of the electron scattering by helium atom%
\begin{equation*}
T_{\mathbf{pp}^{\prime }}\left( z\right) =\left\vert \Psi _{e}(z)\right\vert
^{2}U,
\end{equation*}%
The integration over $\mathbf{p}^{\prime }$ in (\ref{tau}) cancels the
delta-function and one obtains 
\begin{eqnarray*}
\frac{1}{\tau } &=&\int dzN_{He}^{tot}(T,z)\Psi _{e}^{4}(z)\frac{U^{2}m}{%
\hbar ^{3}} \\
&=&\int dzN_{He}^{tot}(T,z)\Psi _{e}^{4}(z)\frac{A\hbar \pi }{m},
\end{eqnarray*}%
where $A=m^{2}U^{2}/\pi \hbar ^{4}$ is the cross section of the electron
scattering on the He atom. Note that the cross section of 2D electrons is
greater than 3D one \cite{Edelman}.

The total density $N_{He}^{tot}(T,z)$ of helium atoms as function of
distance to the surface is a sum of two parts: 
\begin{equation}
N_{He}^{tot}(T,z)=N_{v}(T)+n_{s}(T,z).  \label{ni}
\end{equation}%
The first part $N_{He}^{tot}(T,z)$ is the density of helium vapor. It is
roughly independent of $z$ and is given by 
\begin{equation}
N_{v}=\alpha \left( \frac{Mk_{B}T}{2\pi \hbar ^{2}}\right) ^{3/2}\exp \left( 
\frac{\mu ^{He}-E_{vac}^{He}}{k_{B}T}\right) ,  \label{Ng}
\end{equation}%
where $\alpha =1$ for $^{4}$He and $\alpha =2$ for $^{3}$He because of the
spin degeneracy. The second part $n_{s}(z)$ is the density of SLA. It
depends on the wave function $\Psi _{s}(z)$ of an atom on the surface level: 
\begin{equation}
N_{s}(T,z)=n_{s}(T)\Psi _{s}^{2}(z).  \label{ns}
\end{equation}

Electron mobility 
\begin{equation}
\eta _{e}=\frac{\tau }{m}=\frac{1}{\pi \hbar A\left[
N_{v}(T)I_{v}+n_{s}(T)I_{s}\right] }  \label{mue}
\end{equation}%
where we introduced the notations%
\begin{equation}
I_{s}=\int \Psi _{e}^{4}(z)\Psi _{s}^{2}(z)dz  \label{Is}
\end{equation}%
and%
\begin{equation}
I_{v}=\int \Psi _{e}^{4}(z)dz\approx \int_{0}^{\infty }dz\left[ 2\gamma
^{3/2}z\exp (-\gamma z)\right] ^{4}=3\gamma /8,  \label{Iv}
\end{equation}%
In the absence of the SLA we obtain%
\begin{equation*}
\eta =\frac{8}{3\pi \hbar A\gamma N_{v}(T)}
\end{equation*}%
in agreement with \cite{Saitoh}.

To calculate the integral (\ref{Is}) one need to know the wave function $%
\Psi _{s}^{2}(z)$. An exact calculation of $\Psi _{s}(z)$ is a complicated
many-particle problem. To estimate the contribution of scattering on surface
atoms to the electron mobility one can use an approximate wave function (\ref%
{Psiz}) with $\gamma =\gamma _{s}=\sqrt{-2ME_{SLA}}/\hbar $, where $E_{SLA}$
is the energy of surface level and $M$ is the free atom mass: $%
M^{He4}=6.7\cdot 10^{-24}g$ and $M^{He3}=5.05\cdot 10^{-24}g$. The value $%
E_{s}$ can be taken from the contribution of surface states to the
temperature dependence of surface tension. According to Ref. \cite{SurStates}
$E_{s}^{He4}\approx 3.5K$ and $E_{s}^{He3}\approx 2.25K$, that gives $\gamma
^{He4}\approx (1.3$\AA $)^{-1}$ and $\gamma ^{He3}\approx (1.87$\AA $)^{-1}$%
. In this approximation the integral (\ref{Is}) coming from SLA is

\begin{eqnarray}
I_{s}(T) &=&\int 64\gamma _{e}^{6}\gamma _{s}^{3}z^{6}\exp [-2(\gamma
_{s}+2\gamma _{e})z]dz  \notag \\
&=&\frac{360\gamma _{s}^{3}\gamma _{e}^{6}}{(\gamma _{s}+2\gamma _{e})^{7}}.
\label{Is1}
\end{eqnarray}%
The ratio%
\begin{equation}
r\equiv \frac{n_{s}(T)I_{s}}{N_{v}(T)I_{v}}\approx \frac{n_{s}(T)}{N_{v}(T)}%
\frac{960\gamma _{s}^{3}\gamma _{e}^{5}}{(\gamma _{s}+2\gamma _{e})^{7}}
\label{r}
\end{equation}
may be less or greater than unity depending on temperature and clamping
electric field. This ratio determines the role of SLA in the momentum
relaxation of surface electrons. At low enough temperature, when the vapor
atom density $N_{v}(T)$ is negligible compared to the SLA density because of
the large negative exponent in (\ref{Ng}), the ratio (\ref{r}) is $\gg 1$.
In the opposite limit of high temperature, when $N_{v}(T)$ is not negligibly
small, the ratio (\ref{r}) is $\ll 1$ because of the second factor in the
right hand side of Eq. (\ref{r}), which contains the smallness $\left(
\gamma _{e}/\gamma _{s}\right) ^{4}$. The numerical calculation of the SLA
wave function $\Psi _{s}(z)$ assuming the van der Waals interaction between
He atoms gives a slightly greater value for the ratio (\ref{r}): 
\begin{equation}
r\approx \frac{n_{s}(T)}{N_{v}(T)}\frac{%
2\cdot 10^{4}\gamma _{s}\gamma _{e}^{5}}{(\gamma _{s}+2\gamma _{e})^{5}}.
\label{r1}
\end{equation}

The 2D SLA density $n_{s}(T)$ differs for $^{3}$He and for $^{3}$He.\cite%
{SurStates} For $^{4}$He it is given by density of states of a 2D Bose gas:
with effective mass 
\begin{eqnarray*}
n_{4}(T) &=&\int ~\frac{d^{2}k}{(2\pi \hbar )^{2}}\cdot \frac{1}{\exp \left[
\left( \varepsilon _{4}(k)-\mu _{He4}\right) /T\right] -1} \\
&=&-\frac{M_{4}T}{2\pi \hbar ^{2}}\ln \left[ 1-\exp \left( -\frac{\Delta _{4}%
}{T}\right) \right] ,
\end{eqnarray*}%
where $\Delta _{4}=E_{SLA}^{He4}-\mu ^{He4}\approx 3.6K$ is almost
temperature independent and the SLA effective mass $M_{4}\approx
2.6~M_{0}^{He4}$. For $^{3}$He the analogous calculation would gives 
\begin{eqnarray}
n_{3}(T) &=&2~\int ~\frac{d^{2}k}{(2\pi \hbar )^{2}}\cdot \frac{1}{\exp %
\left[ \left( \varepsilon _{3}(k)-\mu _{He3}\right) /T\right] +1}  \notag \\
&=&\frac{M_{3}T}{\pi \hbar ^{2}}~\ln \left[ 1+\exp \left( -\frac{\Delta
_{3}\left( T\right) }{T}\right) \right] ,  \label{ns3}
\end{eqnarray}%
where%
\begin{equation}
\Delta _{3}\left( T\right) =E_{SLA}^{He3}\left( T\right) -\mu ^{He3}\left(
T\right) .  \label{D3}
\end{equation}%
At $T\rightarrow 0,$ $\Delta _{3}=E_{SLA}^{He3}-\mu ^{He3}\approx 0.25K$,
and $M_{3}\approx 2.25M_{0}^{He3}$. The temperature dependence of $^{3}$He
chemical potential $\mu ^{He3}\left( T\right) $ is stronger than that for $%
^{4}$He and may be essential even at low temperature.\cite%
{DyugJLTP90,DyugSSR90}

To analyze the temperature dependence of the $^{3}$He chemical potential $%
\mu ^{He3}\left( T\right) $ one may apply the exact thermodynamic relation 
\cite{LL5} 
\begin{equation}
\mu =\bar{F}+\frac{P}{n_{L}},  \label{10}
\end{equation}%
where $\bar{F}$ is the free energy per one atom, $P$ is the pressure and $%
n_{L}$ is the liquid density. If temperature is not too high, $P=N_{v}T$,
where the vapor density $N_{v}\left( T\right) $ is given by Eq. (\ref{Ng}).
Since $N_{v}\ll n_{L}$, the second term in Eq. (\ref{10}) is small and the
dependence $\mu _{3}\left( T\right) $ can be found using the relation
between free energy and specific heat $C_{3}(T)$ per one $^{3}$He atom,\cite%
{LL5} 
\begin{equation}
\mu _{3}(T)=\mu
_{3}(0)+\int\limits_{0}^{T}~C_{3}(T_{1})dT_{1}-T~\int\limits_{0}^{T}\frac{%
C_{3}(T_{1})dT_{1}}{T_{1}}.  \label{11}
\end{equation}%
Below we find the temperature dependence $\mu _{3}(T)$ basing on the
experimental data on $C_{3}(T)$.\cite{Greywall} The latter can be fitted
with a high accuracy by the phenomenological formula 
\begin{equation}
C_{3}(T)=0.4T+0.105\frac{T}{T^{2}+T_{0}^{2}};\qquad T_{0}=0.21K,  \label{19}
\end{equation}%
where temperature $T$ is in Kelvins: $[T]=K.$ A theoretical substantiation
of (\ref{19}) is given in Ref. \cite{DyugSSR90}. From (\ref{11},\ref{19})
one obtains: 
\begin{equation}
\Delta _{3}(T)=0.25+0.2T^{2}+0.5T\arctan \left( \frac{T}{T_{0}}\right)
-0.053\ln \left( 1+\frac{T^{2}}{T_{0}^{2}}\right) .  \label{20}
\end{equation}%
Unfortunately, we could not determine from Ref. \cite{ExpMobil} if the
helium vapor density on Fig. 2 of Ref. \cite{ExpMobil} was experimentally
measured or calculated using Eq.\ (\ref{Ng}). In the latter case the
temperature dependence of the chemical potential $\mu _{3}(T)$ taken in Ref. 
\cite{ExpMobil} remains unknown. Therefore, in the comparison of the results
of our calculation with experimental data we take the chemical potential $%
\mu _{3}(T)$ to be temperature independent.

In strong clamping electric field the scattering rate of electrons increases 
due to an increase of the electron velocity in z-direction: 
$v_{ze}\approx \hbar\gamma _{e}/m_{e}$.\cite{Edelman}
For the scattering on vapor atoms this
increase is slower than for the scattering on SLA and ripplons. 
One cannot find analytically the wave
function of surface electrons in the presence of both image potential and 
the clamping field. For an approximate study 
one can apply the variational method with the trial wave
function (\ref{Psiz}), where $\gamma $ is the variational parameter. One gets 
\cite{Shikin} 
\begin{equation}
\gamma =\frac{\gamma _{1}}{3}\left[ \frac{\gamma _{0}}{\gamma _{1}}+\left( 1+%
\sqrt{1-\left( \frac{\gamma _{0}}{\gamma _{1}}\right) ^{6}}\right)
^{1/3}+\left( 1-\sqrt{1-\left( \frac{\gamma _{0}}{\gamma _{1}}\right) ^{6}}%
\right) ^{1/3}\right] ,  \label{gamma}
\end{equation}%
where $\gamma _{1}=(\gamma _{0}^{3}+27\gamma _{F}^{3}/2)^{1/3}$, $\gamma
_{0}\equiv \gamma \left( E_{\perp }=0\right) $, and $\gamma
_{F}^{3}=3meE_{\perp }/2\hbar ^{2}$. Note that $\gamma_e$ enters
 the ratio (\ref{r}),(\ref{r1}) in the fifth
power the ratio (\ref{r}),(\ref{r1}), and in stronger clamping field 
the role of the scattering on SLA
becomes more important. The increase of $\gamma $ given by Eq. (\ref{gamma})
enhances the electron scattering by He atoms and decreases the mobility of
surface electrons. 
For the electron concentration $n_{e}=1.21\cdot
10^{7}cm^{-2}$ as in the experiment on $^{3}$He in Ref. \cite{ExpMobil}
using $eE_{\perp }=2\pi e^{2}n_{e}$ one obtains the value $\gamma
_{F}/\gamma _{0}=0.28$, which gives $\gamma /\gamma _{0}\approx 1.02$. This
ratio changes by only $2\%$ the contribution from He vapor to the electron
scattering rate. This ratio enters in sixth power to the SLA
contribution [see Eqs. (\ref{r}),(\ref{r1})], and according to the above
consideratioon the clamping electric
field at electron density in the experiment \cite{ExpMobil} increases by $%
14\%$ the SLA contribution to the scattering rate of electron on $^{3}$He
surface and by $5\%$ the SLA contribution for $^{4}$He. This correction is
small, though it may increase if more accurate study of the
electron wave function in clamping field on the surface of liquid helium is
performed. We leave this more accurate study for further publications and
disregard this correction in what follows.

To compare the calculated mobility of electrons on $^{4}$He surface with
experiment one has to take into account the electron scattering by ripplons. The
ripplon-limited mobility in weak clamping field is given by \cite{Shikin} 
\begin{equation}
\eta _{R}=\frac{9\sigma \hbar ^{3}}{m^{2}\Lambda ^{2}\gamma ^{2}T}\left[ 
\frac{cm}{dyn\cdot c}\right] ,  \label{muR1}
\end{equation}%
while in strong clamping field $E_{\perp }$ it is 
\begin{equation}
\eta _{R}=\frac{8\sigma \hbar }{m(eE_{\perp })^{2}}.  \label{muR2}
\end{equation}%
The surface tension of $^{4}He$ $\sigma ^{4}=0.37dyn/cm$, and in the limit
of weak clamping field we get 
\begin{equation}
\eta _{R}\approx \frac{3\cdot 10^{7}}{T[K]}\left[ \frac{cm^{2}}{Vc}\right] .
\label{muR1ap}
\end{equation}%
The total electron mobility is roughly given by 
\begin{equation}
\eta _{tot}^{-1}=\eta _{e}^{-1}+\eta _{R}^{-1},  \label{muRtot}
\end{equation}%
where $\eta _{e}^{-1}$ is given by Eq. (\ref{mue}).

From Eq. (\ref{r1}) we obtain for $^{4}$He 
\begin{equation*}
r\approx \frac{5\cdot 10^{-4}}{\sqrt{T}}%
\exp \left( \frac{3.6}{T}\right) .
\end{equation*}%
$r=1$ at $T\approx 0.5K$. Hence, the contribution from the scattering on
SLA becomes greater than that on He vapor at $T\lesssim 0.5K$. At this
temperature the contributions from all three scattering mechanisms are of the
same order. Far from this temperature the contribution from the SLA to the 
electron scattering rate on the surface of $^{4}$He is usually a small 
correction to the total scattering rate. This correction becomes more 
important in strong clamping field, i.e. for higher concentration of 
surface electrons.

At low temperature the $^{3}$He
viscosity becomes very high, and all ripplons disappear in the temperature
region $T_{SF}<T<T_{F}$, where $T_{SF}$ is the transition temperature to the 
$^{3}$He superfluid state. Therefore, in the low temperature part of the phase
diagram the contribution from ripplon to the electron scattering is much weaker
than for He4, making the contribution from SLA more important. 
The results of the calculation of the temperature
dependence of electron mobility on $^{3}$He surface together with the
experimental data from \cite{ExpMobil} are shown in Fig. 1. From Fig. 1 we
see that the agreement between experimental data on electron mobility on the
liquid $^{3}$He surface and the theoretical calculation considerably
improves if the SLA are taken into account. The agreement is better at low
temperature, where the contribution of SLA is stronger than the contribution
from He vapor. However, the inclusion of SLA does explain all discrepancies between theory and
experiment. In particular, the experimental points show a $\sim 2$ times lower
mobility than the theory prediction. This discrepancy may come from the
inaccurate value of the He atom cross section, which enters the electron
scattering rate. 
The crossing of the experimental lines for the
electron mobility $\eta (N_{g})$ as function of He vapor concentration, clearly
seen in Fig. 2 of Ref. \cite{ExpMobil},
also remains unexplained. A little better agreement with experiment can be reached
after taking into account the modification of the surface electron wave
function by clamping electric field [see Eq. (\ref{gamma}) and discussion
after this formula]. 

\begin{figure}[tbh]
\includegraphics{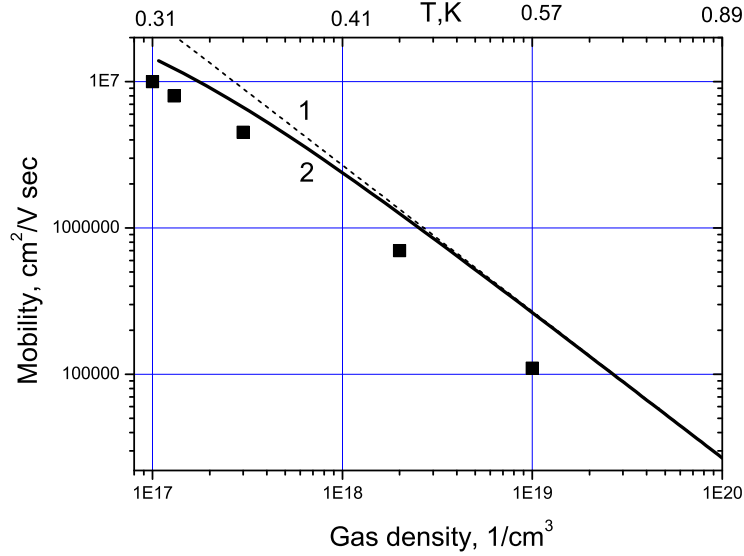}
\caption{The mobility of surface electrons as function of He vapor
density for $^{3}$He in logarithmic scale. The dotes are the experimental
data from \cite{ExpMobil}. The dash line is the theoretical prediction by
Saitoh.\cite{Saitoh} The solid line is our modification of the Saitoh
results where we take into account the SLA contribution to the electron
scattering rate.}
\label{FigMobHe3}
\end{figure}

To summarize, we calculated the contribution to the electron scattering rate
from the surface level atoms, proposed in Ref. \cite{SurStates}. These
SLA explain the long-standing puzzles in the temperature dependence of the
surface tension of both He isotopes and allow to reach a very good agreement 
between theory and experiment.\cite{SurStates} 
These surface atom states also make a contribution to the surface electron
scattering rate and reduce the mobility of the surface electrons. 
This contribution may be essential at low
temperature $T<0.5$ when the He vapor concentration is exponentially small.
The contribution from the SLA improves an agreement between theory and 
experiment
on surface electron mobility. However, this contribution alone is
not sufficient to explain all puzzles in the temperature dependence of the 
surface electron mobility. For more accurate calculation of the electron
mobility one needs to take into account the influence of the clamping
electric field on the surface electron wave function and the temperature
dependence of the He3 chemical potential. 

The work was supported by RFBR grant No 06-02-16551 and by INTAS No 01-0791.

\end{document}